\documentclass[preprint,12pt,3p]{elsarticle}
\usepackage[utf8]{inputenc}
\usepackage[T1]{fontenc}
\usepackage{lmodern}
\usepackage{graphicx,color,framed}
\usepackage{textcomp}
\usepackage{float}
\usepackage{chemfig}
\usepackage{amsmath}
\usepackage{multirow}
\usepackage{microtype}
\usepackage{subcaption}

\begin{document}
\begin{frontmatter}
\title{Carbamazepine solubility in supercritical CO$_2$: a comprehensive study}
\author[isc]{Kalikin N.N. \corref{cor1}}
\ead{nikolaikalikin@gmail.com} \author[isc]{Kurskaya M.V.}
\author[isc]{Ivlev D.V. }
\author[isc]{Krestyaninov M.A.}
\author[isc]{Oparin R.D.} \corref{cor1}
\ead{r.d.oparin@yandex.ru}
\author[lpzg]{Kolesnikov A.L. \corref{cor1}}
\ead{kolesnikov@inc.uni-leipzig.de}
\author[isc,hse]{Budkov Y.A. \corref{cor1}}
\ead{ybudkov@hse.ru}
\author[lille]{Idrissi A.}
\author[isc]{Kiselev M.G. \corref{cor1}}
\ead{mgk@isc-ras.ru}
\address[isc]{G.A. Krestov Institute of Solution Chemistry of the Russian Academy of Sciences, Laboratory of NMR Spectroscopy and Numerical Investigations of Liquids, Akademicheskaya str. 1, 153045, Ivanovo, Russia}
\address[hse]{Tikhonov Moscow Institute of Electronics and Mathematics, School of Applied Mathematics, National Research University Higher School of Economics, 34, Tallinskaya Ulitsa, 123458, Moscow, Russia}
\address[lpzg]{Institut für Nichtklassische Chemie e.V., Permoserstr. 15, 04318, Leipzig, Germany}
\address[lille]{University of Lille, Faculty of Science and Technology, LASIR (UMR CNRS A8516), B\^at. C5, Cit\'e Scientifique, 59655 Villeneuve d'Ascq Cendex, France}
\cortext[cor1]{Corresponding author}

\begin{abstract}
In this paper we present our study of carbamazepine solubility in supercritical carbon dioxide. We have calculated the solubility values along two isochores corresponding to the CO$_2$ densities $\rho = 1.1\rho_{cr}(CO_2)$ and $\rho= 1.3\rho_{cr}(CO_2)$, where $\rho_{cr}(CO_2)$ is the critical density of CO$_2$, in the temperature range from $313$ to $383~K$, as well as along three isotherms at $T=318$, $328$ and $348~K$ by an approach based on the classical density functional theory. The solubility values were also obtained using \textit{in situ} IR spectroscopy and molecular dynamics simulations along the mentioned isochores and isotherms, respectively. Because the density functional theory only takes into account the Lennard-Jones interactions, it can be expected to underestimate the solubility values when compared to the experimental ones. However, we have shown that the data calculated within the classical density functional theory qualitatively reproduce the solubility trends obtained by IR spectroscopy and molecular dynamics simulation. Moreover, the obtained position of the upper crossover pressure is in good agreement with the experimental literature results.
\end{abstract}

\end{frontmatter}

\section{Introduction}
Nowadays a large part of the drugs presented on the pharmaceutical market are classified as Class II, according to the Biopharmaceutics Classification System (BCS) \cite{padrela2018supercritical,yu2002biopharmaceutics,shah2014gl}, i.e. demonstrate high intestinal permeability and low aqueous solubility. Such characteristics can become the reason for refusing from the production of a potentially efficient active pharmaceutical ingredient (API) during its development phase.

A number of approaches have been developed to deal with the poor drug solubility and/or dissolution issues, and at the same time to preserve the pharmacological effect of the compound. They include particle size reduction, formation of amorphous forms, salts, solvates or cocrystals \cite{padrela2018supercritical,healy2017pharmaceutical,bavishi2016spring,serajuddin2007salt,kanaujia2015amorphous}.
Modern micronization techniques utilize specific properties of supercritical fluids, most frequently supercritical carbon dioxide (scCO$_2$) because of its favorable characteristics for process design. These approaches can be divided into three general groups, where scCO$_2$ acts as a solvent, as an antisolvent (co-antisolvent), and as an additive. Each group has its own advantages and disadvantages (Table 6 in \cite{padrela2018supercritical}), and the choice of the method is highly dependent on the compound ability to dissolve in scCO$_2$. Thus, the knowledge of the drug compound solubility in the scCO$_2$ medium is highly valuable for the technological process of the development of a bioactive compound and its potential pharmaceutical applications. However, whether it is an experimental or a computational scheme, most of the conventional approaches, used to obtain information about the solubility of a given drug compound in the supercritical fluid (SCF), are time-consuming and resource-intensive.
Recently \cite{budkov2019possibility}, we have proposed a technique for computing solubility values based on the calculation of the drug molecule solvation free energy in scCO$_2$ within the classical density functional theory (cDFT). There we demonstrated the capability of our approach to qualitatively predict the solubility of ibuprofen in scCO$_2$ and the pressure values, at which we observe changes in the solubility dependence on temperature at a constant pressure value (pressure crossovers).
The calculated solubility values were lower than the experimental ones. The observed discrepancies were attributed to the neglecting of the electrostatic interactions in the framework of the cDFT.

In the present paper, we would like to show the results of our study of carbamazepine (CBZ) solubility in scCO$_2$. CBZ is an essential anticonvulsant and antiepileptic representative \cite{schain1977carbamazepine,wiffen2005carbamazepine}, also used in the treatment of bipolar disorders \cite{kowatch2000effect}. It features on the World Health Organization's List of Essential Medicines \cite{world2014selection} and at the same time belongs to Class II of the BCS \cite{zakeri2009biopharmaceutical}, i.e. demonstrates high permeability and poor solubility in water.  Although there is a fair number of papers featuring CBZ, the solubility data in scCO$_2$ for this compound are practically non-existent in literature (the paper, which we will refer to most is an experimental study by Yamini et al \cite{yamini2001solubilities}). 
The paper is structured as follows: the second section describes the theoretical approach, in the third one we describe the MD simulation details, in the forth - our experimental measurements, and then we discuss the obtained results. 

\section{cDFT-based solubility estimation methodology}
For the theoretical prediction of CBZ solubility we utilized the methodology discussed in depth in our recent papers \cite{budkov2019possibility}. Here we shortly describe it.

We use the following relation to determine the solubility \cite{noroozi2016solvation}, which is based on the equilibrium condition between the solute's solid and solution phases:
\begin{equation}
y_2\approx \frac{p^{sat}}{\rho_b k_BT}\exp(\beta\nu^s[p-p^{sat}]-\beta\Delta G_{solv}),
\label{slblt}
\end{equation}
here $\rho_b$ is the bulk density of scCO$_2$, $k_B$ is the Boltzmann constant, $T$ is the temperature, $p$ is the total pressure imposed in the system, $\beta=(k_BT)^{-1}$; $\nu^s$ and $p^{sat}$ are the molar volume and saturation pressure of the pure solute solid phase, respectively, and $\Delta G_{solv}$ is the solvation Gibbs free energy of the solute molecule in the scCO$_2$ medium. As is seen, one has to know the sublimation parameters of the solute and its solvation free energy to obtain the solubility values. Within our methodology, the data on the solute sublimation, i.e. the molar volume and saturation pressure values, is supposed to be taken from the experimental data available in literature, whereas the solvation free energy is computed with the aid of the cDFT. 

We start the calculation of the solvation Gibbs free energy from writing down the grand thermodynamic potential for the scCO$_2$ fluid in the external potential field with the potential energy $V_{ext}(\mathbf r)$:
\begin{equation}\label{3rd_eq}
  \Omega[\rho(\mathbf r)]=k_BT\int d\mathbf r\rho(\mathbf r)[\ln(\Lambda^3\rho(\mathbf r))-1]+F_{ex}[\rho(\mathbf r)]+\int d\mathbf r \rho(\mathbf r)(V_{ext}(\mathbf r)-\mu),
\end{equation}
where the first contribution is the Helmholtz free energy of the ideal gas, the second one is the excess Helmholtz free energy of the fluid, $\Lambda$ is the thermal de Broglie wavelength, $\mu$ is the chemical potential of the bulk phase at the chosen state parameters.

The CO$_2$ particles are considered as spherically symmetric particles interacting through the effective pairwise Lennard-Jones (LJ) potential with the parameters defined as $\varepsilon_{ff}$ and $\sigma_{ff}$ and the distance of the cut-off $r_c=5\sigma_{ff}$. The potential can be divided into two parts, corresponding to the contributions of the hard-core interactions and the attractive interactions between the molecules of the fluid, at its minimum $r_m=2^{\frac{1}{6}}\sigma_{ff}$ according to the Weeks-Chandler-Andersen (WCA)  procedure \cite{andersen1971relationship}. In this respect, the total excess free energy can be written as follows: 
\begin{equation}
\label{5th_eq}
F_{ex}[\rho(\mathbf r)]=F_{hs}[\rho(\mathbf r)]+F_{att}[\rho(\mathbf r)].
\end{equation}
The hard spheres' contribution (the first term in the right hand side of (\ref{5th_eq})) is determined utilizing Rosenfeld's version of the fundamental measure theory (FMT) \cite{rosenfeld1989free} as follows:
\begin{equation}\label{6th_eq}
F_{hs}[\rho(\mathbf r)]=k_BT\int d\mathbf r\Phi(\mathbf r),
\end{equation}
where $\Phi(\mathbf{r})$ is the excess free energy density (expressed in $k_B T$ units), which is the function of the weighted densities $n_{\alpha}(\mathbf{r})=\int d\mathbf{r}'\rho(\mathbf{r}')\omega^{(\alpha)}(\mathbf{r}-\mathbf{r}')$, which, in their turn, are defined by the weight functions $\omega^{(\alpha)}(\mathbf{r})$. The latter characterize the geometric properties of the hard spheres. Defining expressions for them can be found in the review by R. Roth \cite{roth2010fundamental}. The Barker-Henderson (BH) diameter of the hard sphere is determined within the Pade approximation \cite{Verlet1972a}.

The attractive contribution to the excess free energy is described within the mean-field approximation:
\begin{equation}\label{11th_eq}
  F_{att}[\rho(\mathbf{r})]=\frac{1}{2}\int d\mathbf{r}\rho(\mathbf{r})\int d\mathbf{r}'\rho(\mathbf{r}')\phi_{WCA}(\mathbf{r}-\mathbf{r}'),
\end{equation}
where the effective WCA pair potential of attractive interactions is:
\begin{equation}\label{12th_eq}
  \phi_{WCA}(r) = \left\{
  \begin{array}{lr}
    -\varepsilon_{ff}, & r<r_m \\
    4\varepsilon_{ff}\left[\left(\frac{\sigma_{ff}}{r}\right)^{12}-\left(\frac{\sigma_{ff}}{r}\right)^{6}\right], & r_m<r<r_c.
  \end{array}
  \right.
\end{equation}
The solute molecule is modeled as an external LJ potential:
\begin{equation}
V_{ext}(\mathbf{r})=4 \varepsilon_{sf}\left[\left(\frac{\sigma_{sf}}{r}\right)^{12}-\left(\frac{\sigma_{sf}}{r}\right)^6\right],
\end{equation}
where the effective parameters of the interaction between this molecule and the molecules of the fluid are determined through the Berthelot-Lorenz mixing rules: $\sigma_{sf}=(\sigma_{ss}+\sigma_{ff})/2$ and $\varepsilon_{sf}=\sqrt{\varepsilon_{ss}\varepsilon_{ff}}$. The parameters of the interactions between the two molecules  of the active compound ($\sigma_{ss}$, $\varepsilon_{ss}$) and  two molecules of CO$_2$ ($\sigma_{ff}$, $\varepsilon_{ff}$) can be obtained by fitting the respective parameters of the liquid vapour critical point. One can find the values of the critical temperature and density for CO$_2$ at NIST \cite{nist}. We took the values for the CBZ critical temperature and pressure from the paper by Li et al \cite{li2013new}. The values of all the parameters are presented in Table \ref{table1}.

\begin{table}[h!]
\centering
\caption{The values of the parameters used to calculate the solubility.}
\begin{tabular}{c|c|c|}
\label{table1}
& value & source \\
     \hline
     $\varepsilon_{ff}$   &   $218.73$ $K$                    &   this study \\
     $\sigma_{ff}$        &   $0.336$ $nm$                    &   this study \\
     $\varepsilon_{ss}$   &   $565.91$ $K$                    &   this study \\
     $\sigma_{ss}$        &   $0.718$ $nm $                   &   this study \\
     $\varepsilon_{sf}$   &   $351.83$ $K  $                  &   this study \\
     $\sigma_{sf}$        &   $0.527$ $nm  $                  &   this study \\
     $T_{cr}(CO_2)$       &   $304.13$ $K  $                  &   \cite{nist} \\
     $\rho_{cr}(CO_2)$    &   $10.62$ $mol/l$                 &   \cite{nist} \\
     $T_{cr}(CBZ)$        &   $786.83$ $K$                    &   \cite{li2013new} \\
     $P_{cr}(CBZ)$        &   $25.71$ $bar$                   &   \cite{li2013new} \\
     $\nu(CBZ)$           &   $0.18048$ $m^3/mol$             &   \cite{li2013new}\\
     $p^{sat}(CBZ)$       &   $\ln{p^{sat}}=32.7-13343/T$   &   \cite{drozd2017novel}\\
     
\end{tabular}
\end{table}

To obtain the density profile of the fluid, one has to take a variational derivative of the grand thermodynamic potential with respect to the density and iteratively compute the Euler-Lagrange equation:
\begin{equation}\label{9th_eq}
  \frac{\delta\Omega[\rho(\mathbf{r})]}{\delta\rho(\mathbf{r})}=0,
\end{equation}
which leads to the following expression for density:
\begin{equation}\label{10th_eq}
  \rho(\mathbf{r})=\rho_b\exp\left[\frac{\mu_{ex}(\rho_b,T)-c^{(1)}_{fmt}(\mathbf{r})-\int d\mathbf{r}'\rho(\mathbf{r}')\phi_{WCA}(\mathbf{r}-\mathbf{r}')-V_{ext}(\mathbf{r})}{k_BT}\right],
\end{equation}
where $\mu_{ex}$ is the excess chemical potential of the bulk phase, $c^{(1)}_{fmt}(\mathbf{r})=\delta F_{hs}[\rho(\mathbf{r})]/{\delta\rho(\mathbf{r})}$ is the one--particle direct correlation function of the hard-sphere system within FMT.

Finally, one can compute the solvation Gibbs free energy of the solute molecule in the scCO$_2$ medium as the excess grand thermodynamic potential as follows:
\begin{equation}
  \Delta G_{solv} = \Omega[\rho(\mathbf r)] - \Omega[\rho_b].
\end{equation}

Then, turning back to Eq. (\ref{slblt}), we can obtain the solubility values  after calculating the solvation free energy. The values of the sublimation pressure were obtained, using the empirical relation, presented in the work by Drozd et al \cite{drozd2017novel}, and the molar volume of the compound was taken from the paper by Li et al \cite{li2013new} (Table \ref{table1}).

\section{MD simulation and quantum chemical calculations}
\begin{figure}[h!]
\center{\includegraphics[width=0.7\linewidth]{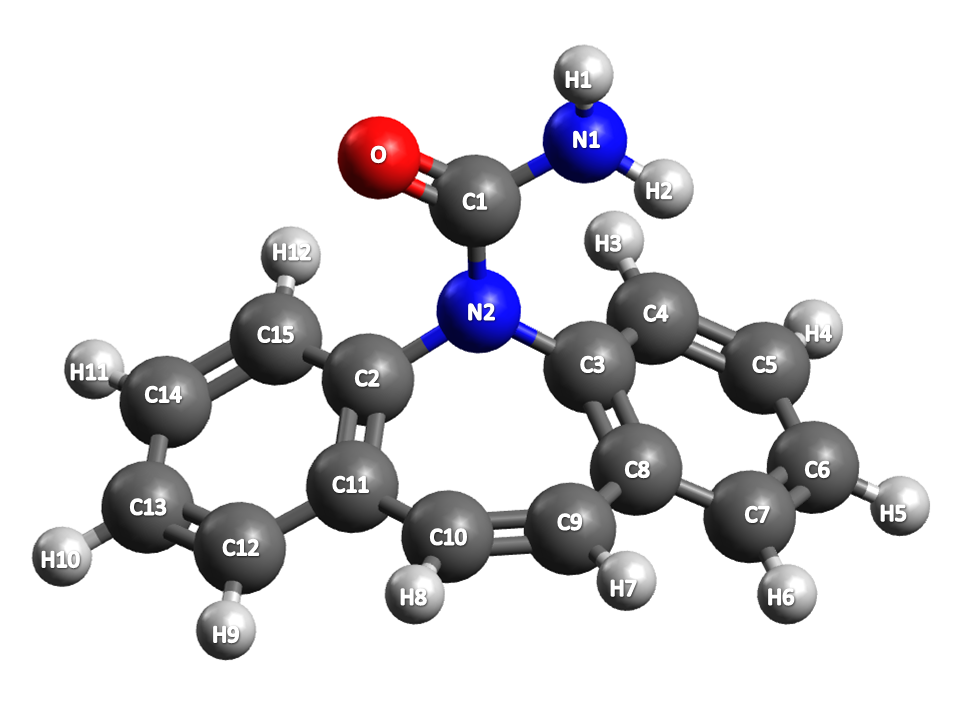}}
\caption{CBZ molecule structure.}
\label{Fig1}
\end{figure}

The CBZ solubility in scCO$_2$ was also estimated by MD simulation,  through the same Eq. (\ref{slblt}) with the same input parameters characterizing the solute sublimation and then by calculating the solvation free energy within the Bennett Acceptance Ratio (BAR) method \cite{bennett1976efficient}.  Such choice was motivated by the fact that, as it was demonstrated \cite{shirts2005comparison,shirts2005solvation}, the BAR method seems to be significantly more effective than the others, when it comes to more realistic modeling tests. Within this method one usually needs to perform a decent number of simulations, each corresponding to a certain intermediate step with its own coupling parameters $\lambda$, resulting in large computational costs \cite{frolov2015accurate}.   Then one obtains the solvation free energy values by the summation of all such transitionary stages from the full solute-solvent interaction coupling to none. We used the following pairwise potential of the interaction between the particles to compute the solvation free energy:
\begin{equation}
\label{eq_MD_pot}
     U_{ij}(\lambda_{LJ},\lambda_C)=\lambda_{LJ}\times4\varepsilon_{ij}\left[\left(\frac{\sigma_{ij}}{r}\right)^{12}-\left(\frac{\sigma_{ij}}{r}\right)^6\right]+\lambda_C\times\frac{q_iq_j}{r},
\end{equation}
where $\lambda_{LJ}$ and $\lambda_C$ are, respectively, the alchemical scaling parameters of Lennard-Jones and Coulombic interactions; $q_i,q_j$ are the atomic charges, $\varepsilon_{ij}$ and $\sigma_{ij}$ are the Lennard-Jones parameters of the solute-solvent interaction. We obtained these parameters through the Berthelot-Lorenz mixing rules: $\sigma_{ij}=(\sigma_{ii}+\sigma_{jj})/2$, $\varepsilon_{ij}=\sqrt{\varepsilon_{ii}\varepsilon_{jj}}$. The simulation was performed in the Gromacs 4.6.7 package \cite{hess2008gromacs}. For each computation of the solvation free energy $12$ independent simulations were performed, each with its own potential energy (Eq. (\ref{eq_MD_pot})), corresponding to a certain pair of the coupling parameters. The following set of the alchemical coefficients $\{\lambda_{LJ},\lambda_{C}\}$ was chosen: $\{0.0,0.0\}$, $\{0.2,0.0\}$, $\{0.5,0.0\}$, $\{1.0,0.0\}$, $\{1.0,0.2\}$, $\{1.0,0.3\}$, $\{1.0,0.4\}$, $\{1.0,0.5\}$, $\{1.0,0.6\}$, $\{1.0,0.7\}$, $\{1.0,0.8\}$, $\{1.0,1.0\}$; here $0.0$ and $1.0$ denote the fully decoupled and fully coupled interactions between the solute and solvent particles, respectively. The structure of the CBZ molecule was taken from the Automated Topology Builder (ATB) \cite{malde2011automated}. We used GROMOS 54A7 force-field \cite{schmid2011definition} for intramolecular CBZ interactions and the Lennard-Jones contribution of intermolecular ones. The atomic partial charges of CBZ were developed based on the Merz-Kollman method \cite{singh1984approach,besler1990atomic}, using Gaussian 09 software \cite{frisch2013gaussian} with the PBE functional and 6-311++g(2d,p) basis set. The atomic partial charges were averaged over two CBZ conformers (Fig.\ref{Fig1}, Table \ref{table2}).

\begin{table}[h!]
\centering
\caption{Calculated partial charges.}
\begin{tabular}{l|l|r|l|l|r}
\label{table2}
      number & atom & charge&number & atom & charge \\
      \hline
1  &   H1	&    0.364805    & 16    & H6     &	 0.148109\\
2  &   N1	&   -0.815994    & 17    & C8     &  0.276173\\
3  &   H2	&    0.330523    & 18    & C9     &	-0.322119\\
4  &   C1	&    0.709618    & 19    & H7     &	 0.152478\\
5  &   O    &   -0.568671    & 20    & C10    & -0.178388\\
6  &   N2	&   -0.305988    & 21    & H8     &	 0.138673\\
7  &   C2   & 	 0.217946    & 22    & C11    &  0.162796\\
8  &   C3   &	 0.070900    & 23    & C12    &	-0.217421\\
9  &   C4   &	-0.130436    & 24    & H9     &	 0.132605\\
10 &   H3   &	 0.137536    & 25    & C13    & -0.116918\\
11 &   C5   &	-0.170959    & 26    & H10    &	 0.126729\\
12 &   H4   &	 0.135166    & 27    & C14    & -0.129340\\
13 &   C6   &	-0.090339    & 28    & H11    &	 0.130257\\
14 &   H5   &	 0.123977    & 29    & C15    &	-0.222715\\
15 &   C7   &	-0.262720    & 30    & H12    &	 0.173718\\
\end{tabular}
\end{table}

The interaction potential for the carbon dioxide molecule corresponds to Zhang's one \cite{zhang2005optimized}. The MD simulation parameters are the same as in the work  by Paliwal and Shirts \cite{paliwal2011benchmark} for the case of the methane solvation free energy calculation, except for the case of the barostat choice, as we used Parinello-Rahman pressure coupling. Each simulation was performed for $100~ps$  in the NVT ensemble with a step of $1~fs$, for $1~ns$ in the NPT ensemble with a step of $2~fs$ and for $10~ns$ of the production run simulation in the NPT ensemble with a step of $2~fs$. We used the experimental data from Yamini's paper \cite{yamini2001solubilities} as the basis and computed two isotherms at $328$ and $348~K$ with pressures corresponding to those presented in the paper.

\section{Infrared spectroscopy experiment}

We also measured experimentally the solubility of CBZ in scCO$_2$ using \textit{in situ} infrared spectroscopy. The obtained values were compared with those obtained by MD and cDFT methods as well as with those obtained by Yamini   \cite{yamini2001solubilities}. 

The experimental solubility values are generally given in the form of  solubility isotherms. However, in our \textit{in situ} IR experiment, we measured the solubility along the isochores, because in the experiment we used a high pressure high temperature (HPHT) optical cell with a constant volume, as it is the case in many supercritical micronization techniques.

In  order to measure the solubility (concentration of a saturated solution) of CBZ in scCO$_2$, we used the approach  developed in our previous works \cite{oparin2016new} \cite{oparin2014dynamic}. This approach is centered around the Beer–Lambert law and, in particular, on the usage of the integral extinction coefficient $\varepsilon_{int}$ (molar absorption coefficient) value of a chosen analytical spectral band. For this purpose, we registered the IR spectra of CBZ, dissolved in an inert solvent that had no specific interactions with the CBZ molecules. The integral extinction coefficient was calculated by the following formula:

\begin{equation}
    \varepsilon_{int}=\frac{A}{lc}
\end{equation}
where $l$ is the optical path length of the sample ($cm$), $c$ is the molar concentration of the solute ($mol\cdot ml^{-1}$), $A$ is the integral intensity ($cm^{-1}$). In its turn, the integral intensity is calculated as follows:
\begin{equation}
    A=\int_{\nu_1}^{\nu_2}\log\frac{I_0(\nu)}{I(\nu)}d\nu,
\end{equation}
where $\nu_1$ and $\nu_2$ are the boundaries of the analytical spectral band, $I_0$ and $I$ are the incident and transmitted intensities, respectively.

\begin{figure}[h!]
\center{\includegraphics[width=1\linewidth]{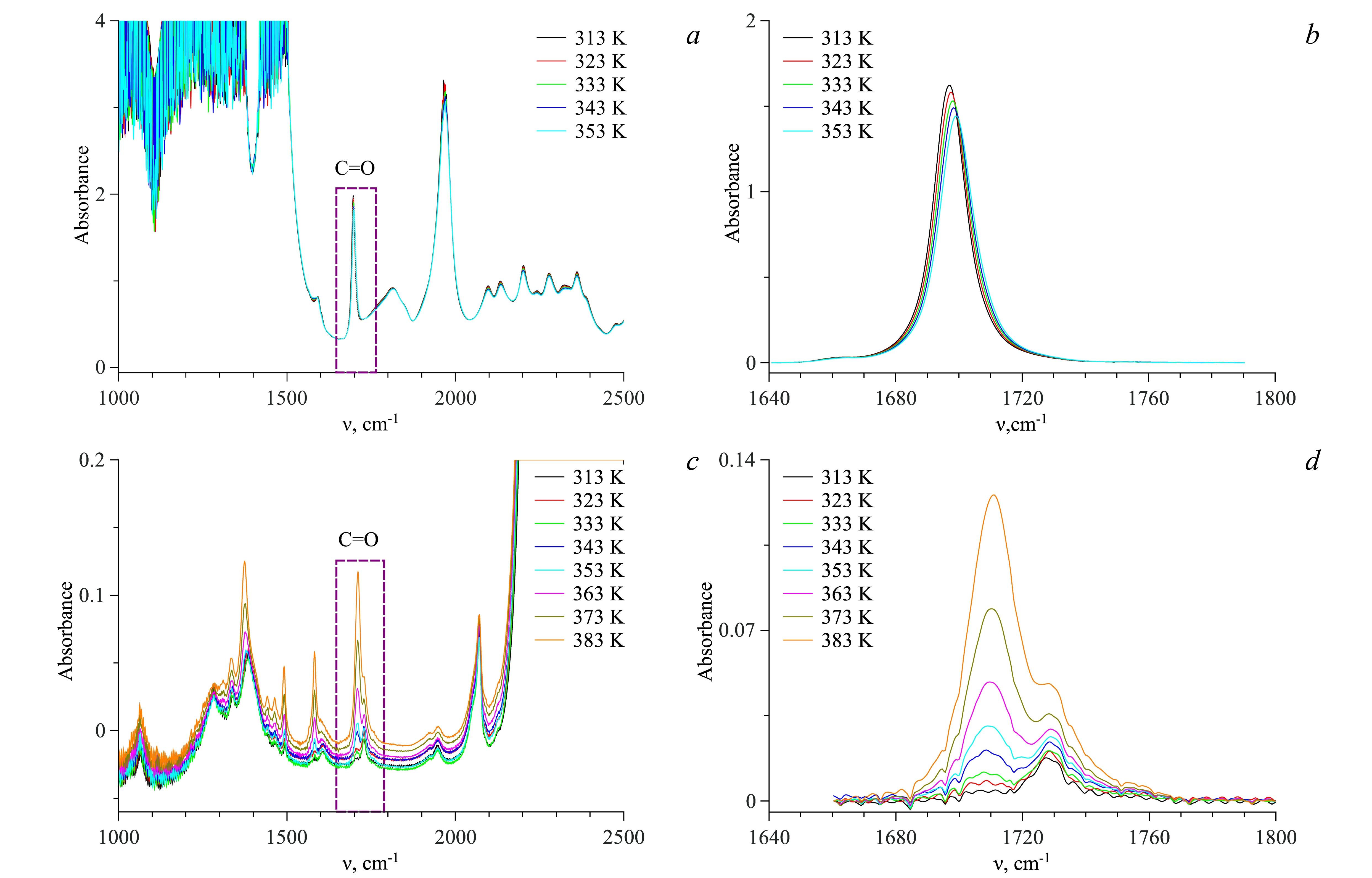}}
\caption{IR spectra of the binary mixtures: CBZ – THF (a), CBZ – scCO$_2$ (c). The spectral domain bounded by a rectangle corresponds to the analytical spectral band of C=O. The analytical spectral band corrected for the base line contribution for CBZ dissolved in THF (b) and in scCO$_2$ (d).}
\label{Fig2}
\end{figure}

Within the \textit{in situ} IR approach, the integral extinction coefficient of the C=O vibration mode of the CBZ chosen as an analytical spectral band was calculated.  This band has a good resolution and high signal–to–noise ratio within the whole studied temperature range.  Then we recorded the IR spectra of the CBZ diluted in tetrahydrofuran (THF) in the temperature range of $313-383$ $K$ with a step of 10 $K$. The molar concentration of the prepared solution was $1.1755\cdot10^{-2}$ $mol\cdot l^{-1}$, which corresponds to the  molar fraction of the CBZ value equal to $9.6012\cdot 10^{-4}$. These spectra were measured on a FTIR spectrometer Bruker VERTEX 80v in the wavenumber range of $1000-4000$ $cm^{-1}$ with a resolution of $1$ $cm^{-1}$ using an HPHT cell \cite{oparin2014dynamic} with the optical path length $l=0.118$ $cm$. In order to increase the spectral signal-to-noise ratio, 128 spectra were recorded for each temperature and then they were averaged out. By subtracting the THF spectra weighted by its corresponding mole faction from the spectra of the CBZ–THF binary mixture, we obtained the CBZ spectrum (Fig. \ref{Fig2}a). The C=O spectral region corrected for the base line is presented in (Fig. \ref{Fig2}b).  

\begin{figure}[h!]
\center{\includegraphics[width=0.7\linewidth]{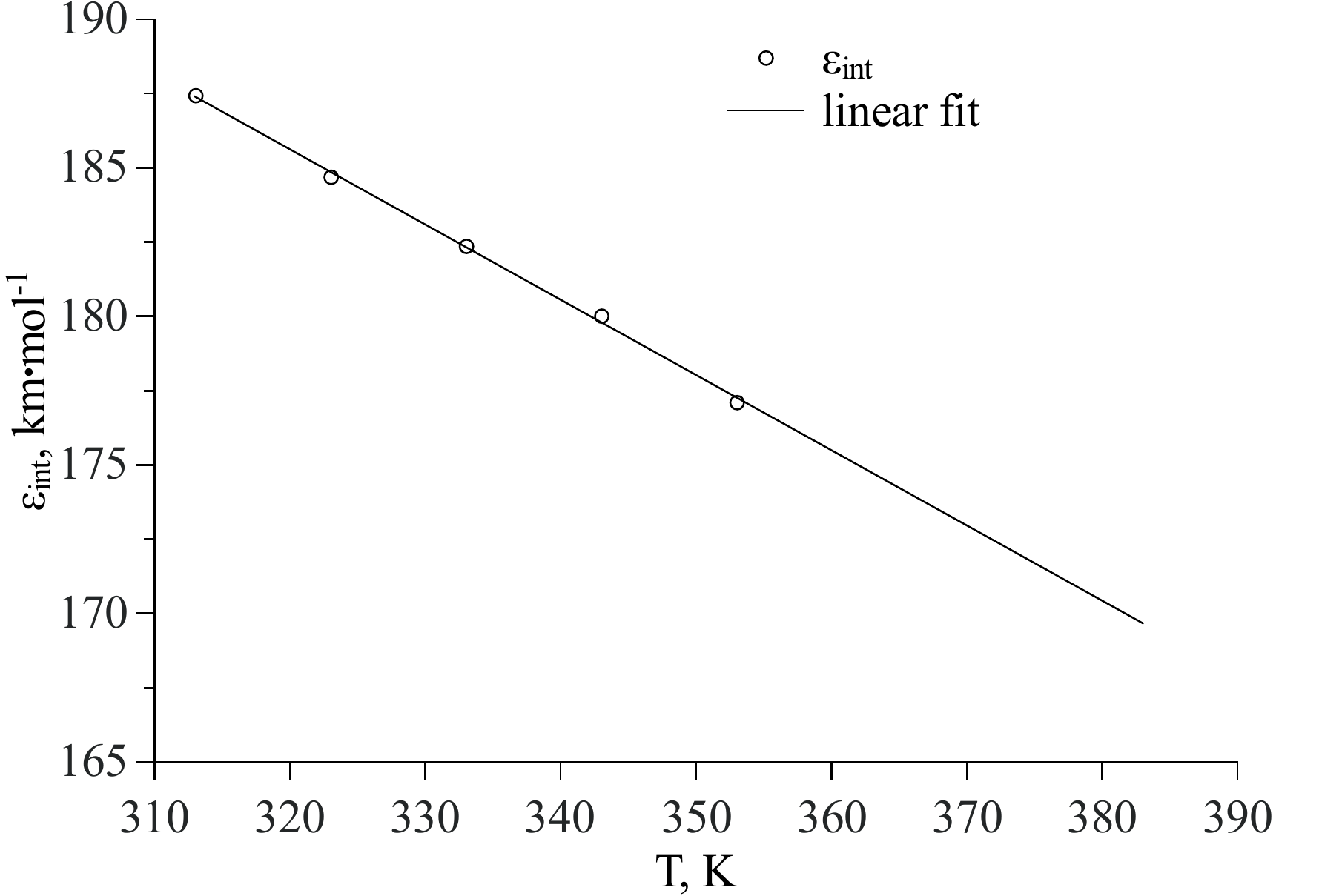}}
\caption{Temperature dependence of integral extinction coefficient and its linear approximation.}
\label{Fig3}
\end{figure}

In order to estimate the value and the standard deviation of the C=O vibration mode integral intensity for each temperature we measured four spectra with a time interval of 10 minutes. The integral extinction coefficients were averaged for each temperature and the standard deviation was found to be from $0.2 \%$ to $0.05 \%$  in the whole studied temperature range. As it is shown in Fig. \ref{Fig3}, the extinction coefficient is described by a linear equation with a small slope coefficient ($\tan\alpha = -0.253$ $km\cdot mol^{-1}\cdot K^{-1}$) at a high accuracy ($R^2=0.998$). The linear character of $\varepsilon_{int}=f(T)$ confirms the absence of specific intermolecular interactions, which was shown in one of our previous works \cite{oparin2005water}.

Within this study, the solubility of CBZ in scCO$_2$ was measured  along two isochores corresponding to the density of the fluid phase $\rho=1.1\rho_{cr}(CO_2)$ and $\rho=1.3\rho_{cr}(CO_2)$ ($\rho_{cr}=10.6249$ $mol/l$ is the critical density of CO$_2$) in the temperature range of $313-383$ $K$ with a step of 10 $K$. The optical path lengths of the sample for these two measurements were  0.108 $cm$ and 0.097 $cm$, respectively.  The C=O stretching region of the IR spectra of CBZ in scCO$_2$, which were measured in the same way as for the CBZ–THF system, was also analyzed (Fig. \ref{Fig2}c,d). These measurements were carried out in a modified high-pressure high-temperature cell described in our work \cite{oparin2019polymorphism}. There was a difference between the shape of the C=O vibration bands in CBZ–THF (Fig. \ref{Fig2}a,b) and CBZ–scCO$_2$ (Fig. \ref{Fig2}c,d) binary systems. In the case of the former system, the C=O vibration mode was symmetric, while in the latter system this band was blue shifted and split into two spectral components, whose intensities and contributions to the total spectral band are sensitive to the temperature increase. We interpret this splitting as a consequence of the presence of two CBZ conformers in the solution. Indeed, in the case of the CBZ–scCO$_2$ system, because of the permanent excess of crystalline CBZ being in contact with the solution phase, there was an equilibrium between the CBZ solid phase and its saturated solution in scCO$_2$. As it was shown \cite{oparin2015interplay,oparin2017screening, oparin2019polymorphism,oparin2020conformational,khodov2014determination, khodov2014inversion}, for drug compounds with different types of polymorphism, there are correlations between the conformers in a solution and polymorphic modifications of their crystalline forms in the bottom phase. Thus, keeping in mind that there is an equilibrium between the solution and the solid  phase, one must expect the presence of a conformational equilibrium of the CBZ molecules in the solution. Indeed, the spectral analysis of the C=O vibration mode in the CBZ–scCO$_2$ system, carried out based on the results of the quantum chemical calculations, showed that there were two conformers in the solution phase that were responsible for the corresponding spectral contributions to the C=O spectral band. A detailed description of the IR experiment and quantum-chemical calculations was presented in our previous papers \cite{ oparin2020conformational,oparin2020correlation}. Therefore, the observed splitting of the analytical spectral band for the CBZ–scCO$_2$ system is a consequence of the presence of two conformers in the solution.

To calculate the concentration of the CBZ in its saturated solution in scCO$_2$ we used the value of integral intensity of the C=O vibration band. To obtain this value, we applied a spectral approximation procedure using Fityk software package \cite{wojdyr2010fityk}. Taking into account the results of the quantum chemical calculations we used two spectral profiles to reproduce the analytical spectral band. Therefore, the full integral intensity of the analytical spectral band was calculated as a sum of these components. Then, the solubility of the CBZ dissolved in scCO$_2$ as a function of temperature was calculated as follows:
\begin{equation}
    c_{CBZ}(T)=\frac{A(T)}{l\cdot\varepsilon_{int}(T)}.
\end{equation}
The molar fraction of CBZ in a scCO$_2$ solution was calculated according to the equation:
\begin{equation}
    X_{CBZ}=\frac{c_{CBZ}}{c_{CBZ}+c_{CO_2}},
\end{equation}
where $c_{CO_2}$ is the CO$_2$ density ($mol\cdot l^{-1}$). 

\section{Results and discussion}
\begin{figure}[h!]
\center{\includegraphics[width=0.7\linewidth]{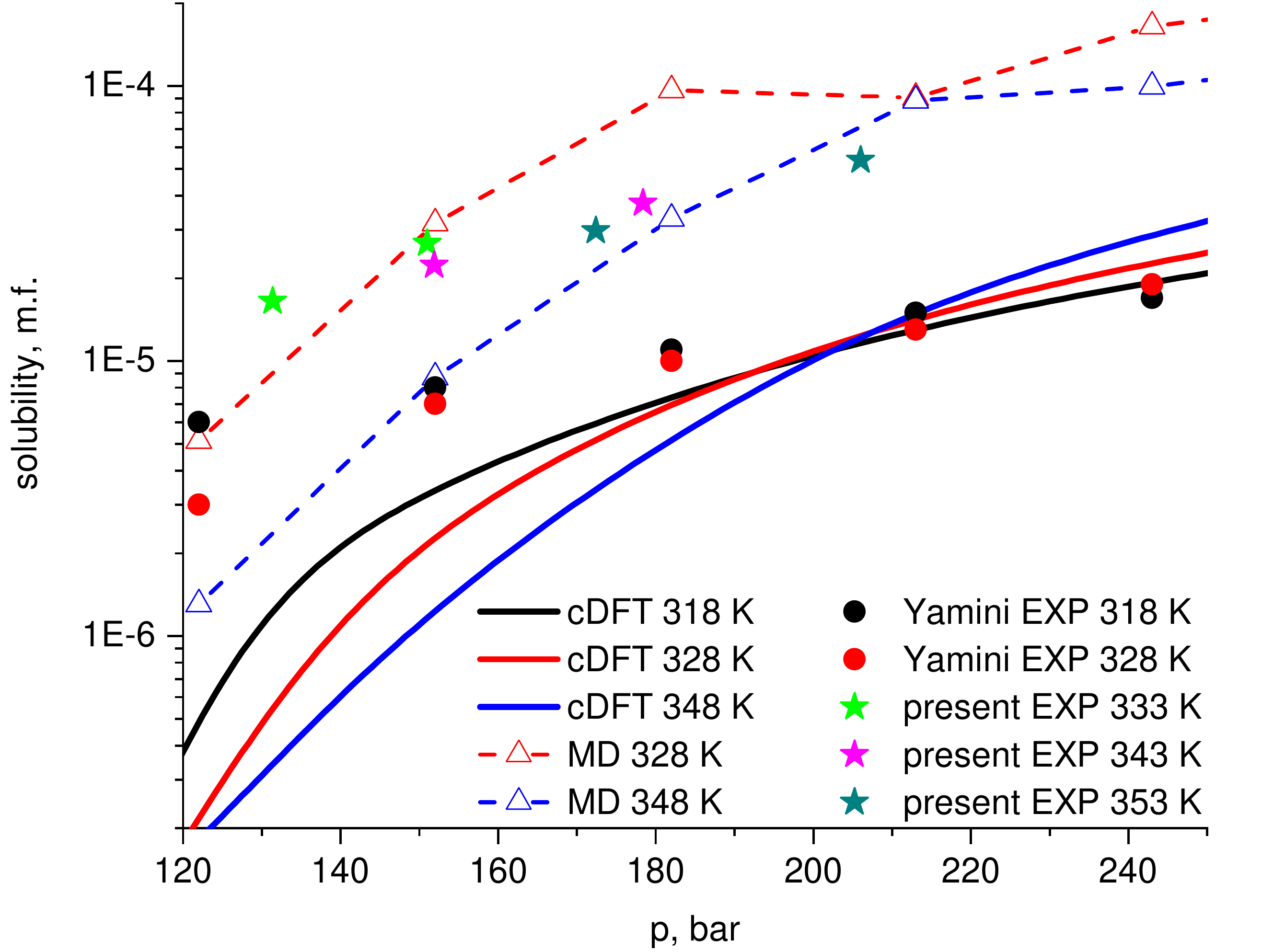}}
\caption{Comparison of the solubility values obtained in accordance with the cDFT approach (solid lines), MD simulation (dashed lines with empty triangles), our \textit{in situ} IR spectroscopy measurements (coloured stars) and the literature data from the experiment by Yamini et al (coloured circles) \cite{yamini2001solubilities}.} 
\label{Fig4}
\end{figure}

Now we compare the computed and experimental solubility values determined in this paper as well as by Yamini et al. \cite{yamini2001solubilities}. The comparison is summarized in Fig. \ref{Fig4}.

As it was outlined above, our \textit{in situ} IR experiment was conducted under the isochoric conditions, thus, it is rather problematic to make a straightforward comparison. Nevertheless, we show the solubility values corresponding to three temperatures $T=333~K$, $T=343~K$ and $T=353~K$, close to the state parameters, at which the values were obtained by the other methods. It should be outlined that the results of the \textit{in situ} IR  experiment correctly represent the solubility behavior within the crossover region, i.e. the inverse dependence of solubility on temperature at a given pressure.

All in all, one can observe qualitative agreement between the obtained results.  
In addition, it is seen that the trend observed in our previous study of the ibuprofen solubility in scCO$_2$ \cite{budkov2019possibility} is the same, i.e. the solubility values obtained according to the cDFT-based approach are lower than the values from the experiments and MD simulations. This can be correlated with the fact that, as we have already mentioned, this method does not take into account the electrostatic contribution to the solvation free energy of the compound in scCO$_2$. 

\begin{figure}[h!]
  \centering
  \begin{subfigure}[b]{0.5\linewidth}
    \centering\includegraphics[width=220pt]{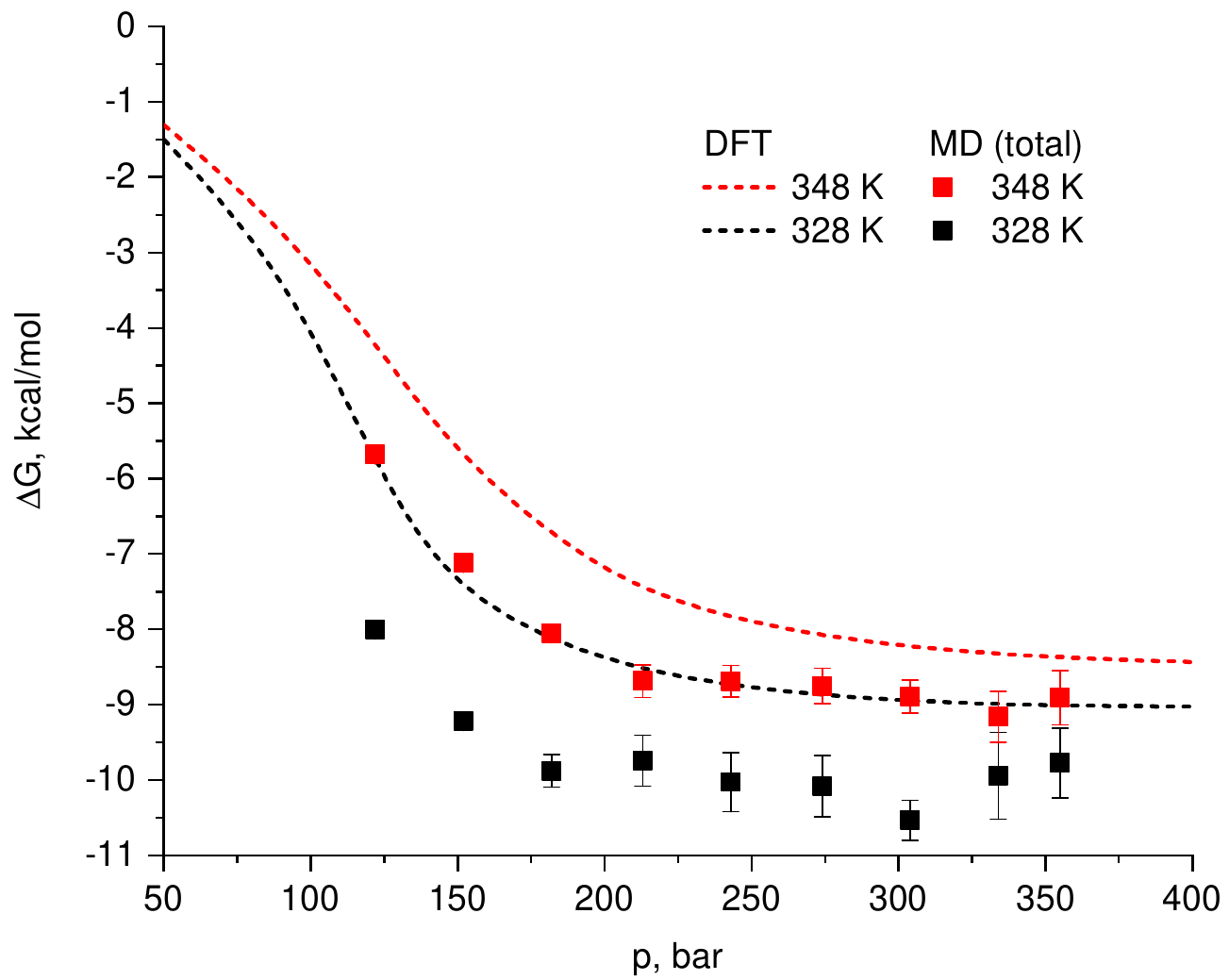}
    \caption{\label{fig:fig1}}
  \end{subfigure}%
  \begin{subfigure}[b]{0.5\linewidth}
    \centering\includegraphics[width=220pt]{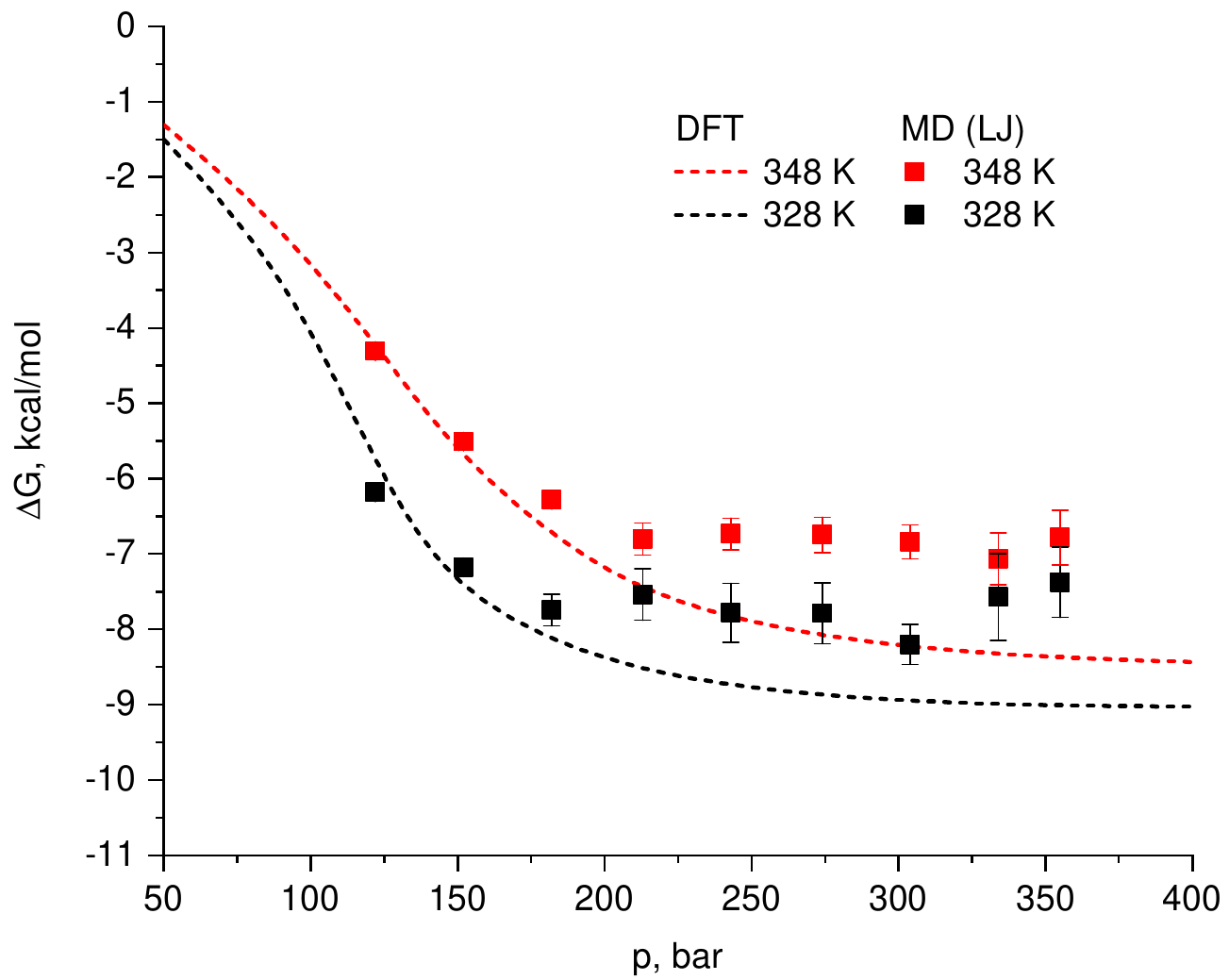}
    \caption{\label{fig:fig2}}
  \end{subfigure}
  \caption{Comparison of the solvation free energy of the compound in scCO$_2$ obtained from cDFT with the total solvation free energy calculated by MD (\subref{fig:fig1}), and with the  Lennard-Jones contribution to the solvation free energy calculated by MD (\subref{fig:fig2}).}
  \label{Fig5}
\end{figure}

It can be seen from the comparison of the solvation free energy values (Fig. \ref{Fig5}a) obtained by the cDFT-based approach (dashed lines) and by the MD simulation (symbols) at two temperatures $T=328~ K$ and $T=348~K$. The discrepancies between the values of the solvation free energy correspond to the neglect of the electrostatic contribution within the cDFT approach. In Fig. \ref{Fig5}b we demonstrate that the values of the cDFT-calculated solvation free energy are in decent agreement with the values of the Lennard-Jones contribution to the solvation free energy obtained by MD simulation, especially within the crossover region, i.e. at the pressures approximately from $120$ to $200$ $bar$.

Despite the underestimation of the solubility magnitude by the cDFT-based approach, it is important to note that the position of the upper pressure crossover is approximately the same for the cDFT, MD simulation and literature results. The reliable value of the lower pressure crossover for CBZ is hard to determine due to its occurrence in the vicinity of the fluid critical point. Then the cDFT-based approach can be utilized as a preliminary stage, used to narrow the working region for the following experimental procedure \cite{foster1991significance}. For instance, the estimation of the location of the solubility crossover region can be helpful in the design of extraction processes based on scCO$_2$ \cite{brennecke1989phase}.

\begin{figure}[h!]
\center{\includegraphics[width=0.7\linewidth]{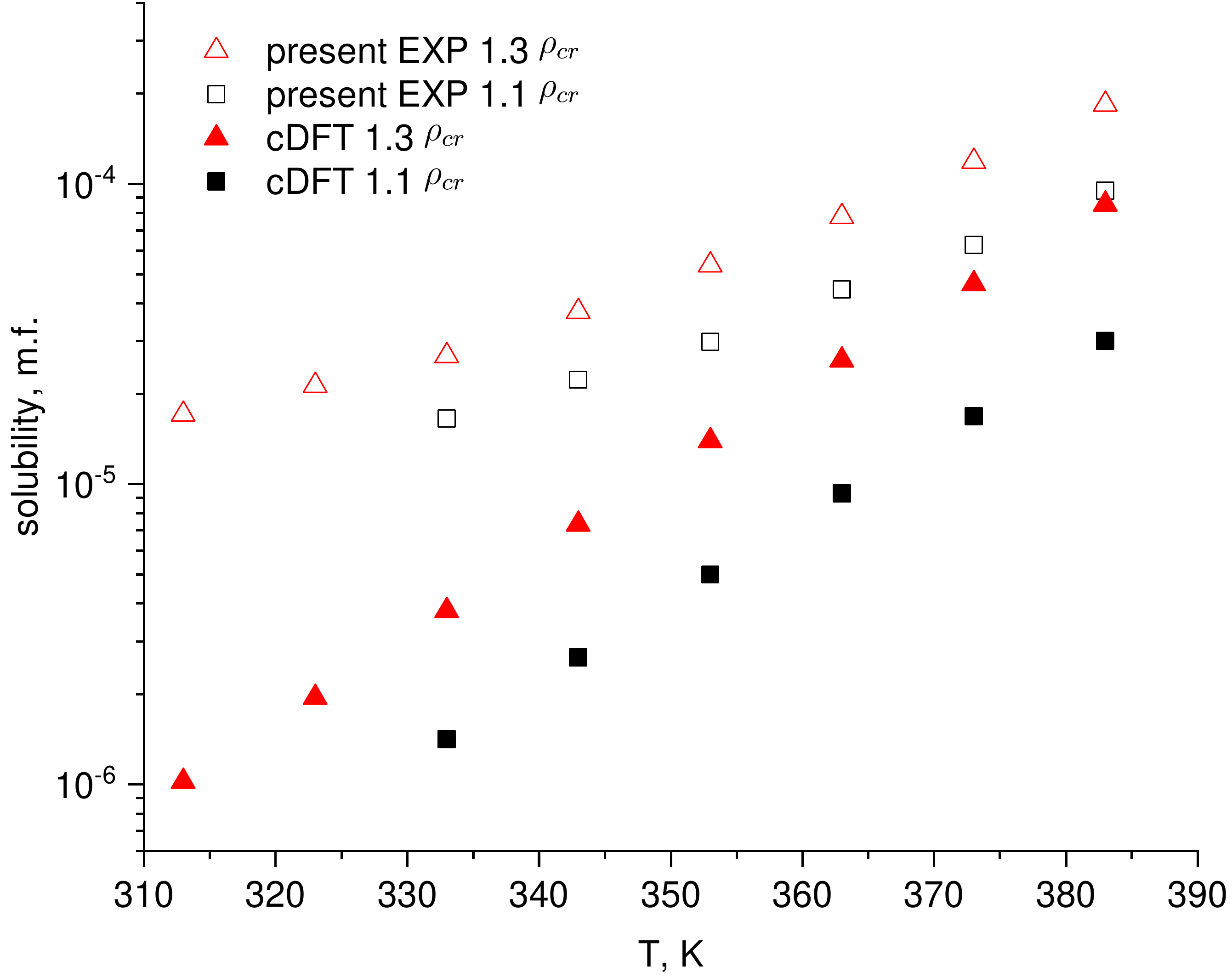}}
\caption{Comparison of the solubility values for two isochores 1.1$\rho_{cr}$ and 1.3$\rho_{cr}$ for CBZ in scCO$_2$ obtained from the cDFT and our \textit{in situ} IR measurements.}
\label{Fig6}
\end{figure}

We compared two isochores, 1.1$\rho_{cr}$ and 1.3$\rho_{cr}$ obtained from our \textit{in situ} IR measurements with the ones computed in accordance with the cDFT approach. The values of the CBZ concentration in its saturated solution in scCO$_2$, as obtained from the experimental measurements, are shown in Table \ref{table3}. In Fig. \ref{Fig6} we show a comparison of the solubility data for these isochores (the coloured symbols are the theoretical results, the empty symbols are the \textit{in situ} IR experiment). As it is expected, the outcomes of the cDFT method underestimate solubility values. The discrepancies in the data are of an order of magnitude for the low temperature range, but the increase in the temperature leads to a decrease in this difference. This is due to the fact that the electrostatics contribution decreases with the temperature increase.

\begin{table}[h!]
\centering
\caption{The temperature dependence of the CBZ concentration in the saturated solution in scCO$_2$ and its molar fraction in scCO$_2$ for two isochores.}
\begin{tabular}{c|c|c|c|c|c|c}
\label{table3}
      & \multicolumn{3}{c}{Isochore 1.1} & \multicolumn{3}{|c}{Isochore 1.3} \\
      \hline
     $T$, $K$ & $p$, $bar$ & $c\cdot10^{4}$, $mol\cdot l^{-1}$ & $X\cdot 10^{5}$, $m.f.$ & $p$, $bar$ & $c\cdot10^{4}$, $mol\cdot l^{-1}$ & $X\cdot 10^{5}$, $m.f.$\\
     \hline
     313  &        &       &      & 97.54  & 2.36  & 1.71  \\
     323  &        &       &      & 124.00 & 2.94  & 2.13  \\
     333  & 131.43 & 1.93  & 1.65 & 151.02 & 3.71  & 2.68  \\
     343  & 151.89 & 2.61  & 2.23 & 178.41 & 5.19  & 3.76  \\
     353  & 172.43 & 3.48  & 2.98 & 206.02 & 7.43  & 5.37  \\
     363  & 193.02 & 4.96  & 4.25 & 233.76 & 10.75 & 7.78  \\
     373 & 213.63 & 7.33  & 6.27 & 261.56 & 16.39 & 11.87  \\
     383 & 234.25 & 11.09 & 9.49 & 298.41 & 25.43 & 18.41  \\
\end{tabular}
\end{table}

We would also like to discuss the potential limitations and restrictions of the proposed cDFT-based approach.
As the input parameters we need to define the critical temperature and pressure of the bioactive compound to obtain the parameters of the LJ interaction potential, and also find the data for the sublimation pressure and molar volume of the compound, as it was discussed above. Such task may have some pitfalls. It is clear that the critical parameters for the API molecule can be determined only through some approximations. In our case the values of the critical parameters for CBZ taken from the literature were calculated within the Group Contribution Method \cite{klincewicz1984estimation}. In short, such method relates, in a certain way, the thermodynamic properties of a compound with its molecular structure. The thing is that there are a number of such correlations, which can in the end lead to a drastic divergence, with CBZ being no exception. For example, the value of the critical temperature is $1197.01~K$ (Kikic et al \cite{kikic2010solubility}) and $786.83~K$ (Li et al \cite{li2013new}). In principle, one can try to fit the
experimental data of solubility using the potential parameters, but in this case cDFT would fail to estimate the position of the crossover points. We believe that the possibility of correct estimation of the pressure crossover is crucial for further applications. Thus, one has to bear in mind that inaccurate estimation of the compound critical parameters can lead to incorrect results. Another problem is the accurate determination of the sublimation pressure values and molar volume of the compound as they also play a crucial role in the determination of the solubility values. Again, approximate methods do not always seem to produce decent results \cite{li2013new}, as compared with the strict experiment \cite{drozd2017novel}.

\section{Conclusions and prospects}
We have obtained values of CBZ solubility in scCO$_2$ using the experimental \textit{in situ} IR approach, MD simulation and theoretical computation. All the obtained results were also compared with the experimental data available in literature.

We have measured two solubility isochores corresponding to the density of $1.1\rho_{cr}$ and $1.3\rho_{cr}$ via the \textit{in situ} IR spectroscopy, where the extinction coefficient of the C=O vibration was first determined in a mixture of CBZ and THF at infinite dilution and at different temperatures in the range between $313~K$ and $353~K$. This allows an accurate determination of the solubility using the Beer-Lambert law. We have computed the solvation free energy of CBZ in scCO$_2$ within the MD simulation for two isotherms: $328~K$ and $348~K$, and calculated the solubility based on these results. The same isochores and isotherms were computed according to the theoretical cDFT-based approach. The results of the latter are in qualitative agreement with the experiment and simulation, and, despite the neglect of the electrostatic solute-solvent interactions, they reproduce the position of the upper pressure crossover with reasonable accuracy.

In conclusion, we would like to speculate on the possible applications of our cDFT-based approach. Firstly, this approach could be utilized for predicting the pressure crossover region within the region of the active compound solubility in scCO$_2$ provided that the sublimation pressure and the critical parameters for the solute are available. In addition, this approach can be implemented as a software tool in a possible complex experimental setup of solubility data treatment along with the \textit{in situ} IR-based methodology presented above.

\section{Acknowledgements}
The research was supported by the Russian Federal Program (grant no. RFMEFI61618$\times$0097). The IR spectroscopy experiment was performed using the molecular fluid spectroscopy facility
(http://www.ckp-rf.ru/usu/503933/) of G.A. Krestov Institute of Solution Chemistry of the Russian Academy of Sciences (ISC RAS) (Russia). MD simulations were performed using the supercomputer facilities provided by NRU HSE.

\newpage
\bibliographystyle{ieeetr}
\bibliography{literature}

\end{document}